\RequirePackage{fix-cm}
\RequirePackage{amsmath}
\documentclass[twocolumn,final]{svjour3}          
\smartqed  
\usepackage{floatrow}
\usepackage{amsfonts}
\usepackage{amssymb}
\usepackage{graphicx}
\usepackage{psfrag}
\usepackage{subfigure}
\usepackage{subfig}
\usepackage{array}
\usepackage{eurosym}
\usepackage{setspace}
\usepackage{algorithm}
\usepackage{algorithmic}
\usepackage{tikz}
\usepackage{bm}
\usepackage[numbers,sort&compress]{natbib}
\usepackage{extarrows}
\usepackage{fancynum}
\usepackage{color}
\usepackage{caption}
\setfnumgsym{\allowbreak}
\DeclareCaptionSubType[alph]{figure}
\journalname{Journal of Signal Processing Systems}
\begin{document}

\title{Residual-Based Detections and Unified Architecture for Massive MIMO Uplink
}


\newtheorem{Thm}{Theorem}
\newtheorem{Lem}{Lemma}
\newtheorem{Cor}{Corollary}
\newtheorem{Def}{Definition}
\newtheorem{Exam}{Example}
\newtheorem{Alg}{Algorithm}
\newtheorem{Prob}{Problem}
\newtheorem{Rem}{Remark}
\renewcommand\thesubsection{\thesection.\Alph{subsection}}
\renewcommand{\algorithmicrequire}{\textbf{Input:}}
\renewcommand{\algorithmicensure}{\textbf{Output:}}

\author{Chuan Zhang \and Yufeng Yang \and Shunqing Zhang \and Zaichen Zhang \and Xiaohu You 
}


\institute{Chuan Zhang$^{\sharp,*}$ \and Yufeng Yang$^{\sharp}$ \and Zichen Zhang \and Xiaohu You \at
              Lab of Efficient Architectures for Digital-communication and Signal-processing (LEADS), National Mobile Communications Research Laboratory, Quantum Information Center, Southeast University, Nanjing, China \\
              \email{\{chzhang, yfyang, zczhang, xhyu\}@seu.edu.cn} \\
              $^{\sharp}$contributed equally to this work, $^*$corresponding author
           \and
           Shunqing Zhang \at
           Shanghai Institute for Advanced Communications and Data Science, Shanghai University, Shanghai, China.\\
           \email{shunqing@shu.edu.cn}}

\date{Received: March 2, 2018 / Accepted: date}

\maketitle

\begin{abstract}
Massive multiple-input multiple-output (M-MIMO) technique brings better energy efficiency and coverage but higher computational complexity than small-scale MIMO. For linear detections such as minimum mean square error (MMSE), prohibitive complexity lies in solving large-scale linear equations. For a better trade-off between bit-error-rate (BER) performance and computational complexity, iterative linear algorithms like conjugate gradient (CG) have been applied and have shown their feasibility in recent years. In this paper, residual-based detection (RBD) algorithms are proposed for M-MIMO detection, including minimal residual (MINRES) algorithm, generalized minimal residual (GMRES) algorithm, and conjugate residual (CR) algorithm. RBD algorithms focus on the minimization of residual norm per iteration, whereas most existing algorithms focus on the approximation of exact signal. Numerical results have shown that, for $64$-QAM $128\times 8$ MIMO, RBD algorithms are only $0.13$ dB away from the exact matrix inversion method when BER$=10^{-4}$. Stability of RBD algorithms has also been verified in various correlation conditions. Complexity comparison has shown that, CR algorithm require $87\%$ less complexity than the traditional method for $128\times 60$ MIMO. The unified hardware architecture is proposed with flexibility, which guarantees a low-complexity implementation for a family of RBD M-MIMO detectors.

\end{abstract}

\section{Introduction}\label{sec:intro}
Multiple-input multiple-output (MIMO) is a key technique for wireless communications \cite{larsson2014massive} and has been incorporated into standards such as the $3$rd generation partnership project ($3$GPP) long term evolution (LTE) and IEEE $802.11$n \cite{lee2009mimo}. By equipping hundreds of antennas at transmitters and serving relatively a small number of users \cite{huh2012achieving}, its advanced version massive MIMO (M-MIMO) provides significant improvement in spectral efficiency, interference reduction, transmit-power efficiency, and link reliability \cite{lu2014overview}.

Because of the large antenna number at base station (BS) or user side, computational complexity becomes unaffordable in M-MIMO detection. Among existing detections, zero forcing (ZF) is a basic way, which neglects the effect of noise \cite{spencer2004zero}. However, its performance is not satisfactory. Though linear schemes like minimum mean square error (MMSE) \cite{larsson2009mimo} improve the performance compared with ZF, its computation complexity still increases drastically as the antenna number grows. For a M-MIMO channel $\mathbf{H}$, computational complexity of MMSE inversion is $\mathcal{O}(M^{3})$, which makes it costly in applications \cite{krishnamoorthy2013matrix}. To avoid matrix inversion, \emph{Neumann} series expansion (NSE) \cite{wang2015efficient,liang2015coefficient,wu2014large} has been employed for approximation. However, complexity remains unaffordable when NSE terms become more than $2$. Thus, iterative linear solvers are proposed for further reduction, such as Gauss-Seidel \cite{dai2015low,wu2016efficient} and conjugate gradient (CG) \cite{yin2015vlsi,yin2014conjugate}. Methods like successive over-relaxation (SOR) \cite{zhang2016large,gao2014matrix} and its variation \cite{yu2016efficient} are also considered. Meanwhile, efficient optimizations of algorithms are also proposed like precondition \cite{xue2016fast,jin2017split}. Most iterative linear detectors can reduce MMSE's complexity to $\mathcal{O}(M^{2})$ with tolerable performance loss.

It is worth noting that existing algorithms mainly focus on approximating exact solution \cite{yang2015fifty}, whereas this paper proposes residual-based detection (RBD) algorithms which focus on the minimization of residual norm. Firstly, minimal residual (MINRES) algorithm, which is a basic RBD, is considered. Its extended version, generalized minimal residual (GMRES) by \cite{abdaoui2007gmres} is also considered, with unavoidable drawbacks, which will be detailed below. In M-MIMO scenario, GMRES can be derived into another version: conjugate residual (CR). Computation process and convergence proof of each RBD algorithm are elaborated. Numerical results under different antenna configurations and correlations are given as well. For lower complexity, the iteration number is chosen as $2$, $3$, and $4$, respectively. Complexity comparison among proposed RBD algorithms and the traditional one is also shown, to demonstrate RBD algorithms' advantages in performance and complexity.

For application considerations, efficient hardware architectures of RBD algorithms are required. In this paper, hardware architectures of MINRES and CR algorithms are proposed. However, as a family of algorithm, computation similarity can be referred and accordingly a unified design method is proposed. The hardware architecture can be given by two common modules: iterative module and coefficient module. Unified hardware architecture for both MINRES and CR algorithms is further proposed, which can also take care of GMRES. Moreover, the proposed design method can be also applied in some RBD and other iterative detectors.

The remainder of the paper is organized as follows. Section~\ref{sec:system_model} gives the system models of non- and correlated M-MIMO detectors. Section~\ref{sec:RBD} employs RBD algorithms and shows the convergence proof for each algorithm. Numerical results are given in Section~\ref{sec:Numerical}. Section~\ref{sec:complexity} elaborates the computational complexity of RBD algorithms. Section~\ref{sec:hardware} proposes the hardware architectures of RBD algorithms and the unified design method. Finally, Section~\ref{sec:con} concludes the entire paper.

\textbf{Notation:} The lowercase and upper bold face letters stand for column vector and matrix, respectively. The operations $(.)^{T}$ and $(.)^{H}$ denote transpose and conjugate transpose, respectively. The entry in the $i$-th row and $j$-th column of $\mathbf{A}$ is $\mathbf{A}(i,j)$. The vector $\boldsymbol\alpha$ in the $k$-th iteration is $\boldsymbol\alpha_{k}$. Complexity is denoted in terms of complex-valued multiplication number.

\section{System Model for Massive MIMO Uplink}\label{sec:system_model}
\subsection{Linear Detection Model}
Consider an uplink of a massive MIMO system with $N$ antennas at the base station (BS), which simultaneously serves $M$ single antenna users. Here, $N$ is always much bigger than $M$ ($N>> M$). The transmitted signal and received vectors are denoted by $\mathbf{s}=[s_{1},s_{2},...,s_{M}]^T $ and $\mathbf{y}=[y_{1},y_{2},...,y_{N}]^T $, respectively, where $\mathbf{s}\in\mathbb{C}^{M}$, $\mathbf{y}\in\mathbb{C}^{N}$ . Then the system model is
\begin{eqnarray}\label{eqn:syst_model}
\mathbf{y} = \mathbf{H} \mathbf{s} + \mathbf{n},
\end{eqnarray}
where $\mathbf{H}$ is an $N \times M$ uplink channel matrix, $\mathbf{n}$ is the vector representing Additive White \emph{Gaussian} Noise (AWGN) with zero-mean and variance $\sigma^{2}$.

According to MMSE equalization scheme, at the BS side, the estimate of the transmitted symbol vector $\mathbf{\hat{s}}$ is
\begin{eqnarray}\label{eqn:two}
\mathbf{\hat{s}}= (\mathbf{H}^H\mathbf{H}+{\sigma}^2\mathbf{I}_{M})^{-1}\mathbf{H}^H\mathbf{y}= \mathbf{A}^{-1}\mathbf{\tilde{y}},
\end{eqnarray}
where the matrix $\mathbf{I}$ means identity matrix with dimension $M$, and the MMSE filtering matrix $\mathbf{A}$ is defined based on \emph{Gram} matrix $\mathbf{G}$:
\begin{eqnarray}\label{eqn:three}
\mathbf{A}=\mathbf{G}+{\sigma}^2\mathbf{I}_{M},
\end{eqnarray}
where $\mathbf{G}=\mathbf{H}^{H}\mathbf{H}$.

Correspondingly, output of matched filter $\mathbf{\tilde{y}}$ is
\begin{eqnarray}
\mathbf{\tilde{y}}=\mathbf{H}^H\mathbf{y}.
\end{eqnarray}

Nevertheless, computational complexity of exact matrix inversion $\mathbf{A}^{-1}$ is $\mathcal{O}(M^{3})$. Methods such as \emph{Cholesky} decomposition based method are not suitable for M-MIMO detection when its scale increases.

\subsection{Correlated Channel Model}\label{sec:channel_model}
Consider correlation of antennas for M-MIMO, this paper applies \emph{Kronecker} model in \cite{kermoal2002stochastic} and $\mathbf{H}$ can be denoted by $\mathbf{H}=\mathbf{R}_{r}^{1/2}\mathbf{W}\mathbf{R}_{t}^{1/2}$, where $\mathbf{W}\in\mathbb{C}^{N\times M}$ is an $N\times M$ i.i.d. channel matrix with zero mean and unit variance. Meanwhile $\mathbf{R}_{r}\in\mathbb{C}^{N\times N}$ and $\mathbf{R}_{t}\in\mathbb{C}^{M\times M}$ are spatial correlation matrices at BS and user side:
\begin{eqnarray}\label{eqn:R_r}
\mathbf{R}_{r}(i,k)=
\begin{cases}
(\zeta_{r}e^{j\theta})^{k-i}, &i\leq k,\\
\mathbf{R}_{r}^{'}(k,i), &i>k;\\
\end{cases}
\end{eqnarray}

\begin{eqnarray}\label{eqn:R_t}
\mathbf{R}_{t}(i,k)=
\begin{cases}
(\zeta_{t}e^{j\theta})^{k-i}, &i\leq k,\\
\mathbf{R}_{t}^{'}(k,i), &i>k.\\
\end{cases}
\end{eqnarray}

The $i$-th row and $k$-th column is denoted by $\mathbf{R}(i,k)$. $\mathbf{R}_{t}^{'}$ and $\mathbf{R}_{r}^{'}$ are conjugate matrices of $\mathbf{R}_{t}$ and $\mathbf{R}_{r}$, respectively. This paper contains four scenarios of correlation condition to elaborate common M-MIMO detectors.
\begin{itemize}
\item[-] \emph{Uncorrelated}: In this condition, correlations of BS and users are ignored, which means correlation factor $\zeta_{t}=\zeta_{r}=0$. Under this circumstance, $\mathbf{R}_{t}^{1/2}$ and $\mathbf{R}_{r}^{1/2}$ are actually $\mathbf{I}_{N}$ and $\mathbf{I}_{M}$, respectively. Then $\mathbf{H}$ is the ideal i.i.d. \emph{Rayleigh} fading channel matrix.
\item[-] \emph{User Correlated}: For multi-antenna users, if the distance between two BS antennas is larger than half-wavelength, correlation between BS antennas can be neglected. In this condition $\mathbf{R}_{r}^{1/2}$ becomes diagonal matrix $\mathbf{D}_{r}$ thus $\mathbf{H}=\mathbf{D}_{r}\mathbf{W}\mathbf{R}_{t}^{1/2}$.
\item[-] \emph{BS Correlated}: For single-antenna users, correlation among users is omitted. Nevertheless, as M-MIMO contains large-scale antenna array, pathloss between BS and users cannot be ignored. Thus the channel is $\mathbf{H}=\mathbf{R}_{r}^{1/2}\mathbf{W}\mathbf{D}_{t}$, where $\mathbf{D}_{t}$ is a diagonal matrix where pathloss attenuation factor is represented.
\item[-] \emph{Fully Correlated}: When fully correlated, both user and BS should be considered. Thus matrix remains $\mathbf{H}=\mathbf{R}_{r}^{1/2}\mathbf{W}\mathbf{R}_{t}^{1/2}$, where $\mathbf{R}_{r}^{1/2}$ and $\mathbf{R}_{t}^{1/2}$ are shown in Eq.s~(\ref{eqn:R_r}) and~(\ref{eqn:R_t}), respectively.
\end{itemize}

\section{Residual-Based Detection Algorithms}\label{sec:RBD}
In this section residual-based detection (RBD) is proposed as a series. For a linear detection problem
\begin{equation}
\mathbf{A}\mathbf{s}=\mathbf{y},
\end{equation}
suppose that $\mathbf{s}_{*}$ denotes the exact estimation of detection signal, existing detection methods mainly focus on the approximation of $\mathbf{s}$ to $\mathbf{s}_{*}$, which is denoted by the absolute error $\|\mathbf{s}-\mathbf{s}_{*}\|$. Whereas vector $\mathbf{r}=\|\mathbf{y}-\mathbf{A}\mathbf{s}\|$ denotes the residual norm of the signal, RBD algorithms mainly focus on the minimization of vector $\mathbf{r}$ in the computation process. This section will give detailed description of RBD algorithms and the relationship between these algorithms will be given too.

\subsection{Minimal Residual Algorithm}\label{sec:MINRES}
As a kind of projection algorithm for massive MIMO detection, proposed minimal residual (MINRES) \cite{saad2003iterative} is the simplest algorithm for its short calculation process, which is shown in \emph{Algorithm}~\ref{MR}.
\begin{algorithm}
\begin{algorithmic}[1]
\caption{Minimal Residual Algorithm}
\label{MR}
\REQUIRE $\mathbf{A}$ and $\mathbf{\tilde{y}}$
\FOR{$k=0,\ldots,K$}
\STATE$\mathbf{r}_{k}=\mathbf{\tilde{y}}-\mathbf{A}\mathbf{s}_{k}$
\STATE$\boldsymbol{\alpha}_{k}=\frac{\mathbf{r}_{k}^{H}\mathbf{A}\mathbf{r}_{k}}{\|\mathbf{A}\mathbf{r}_{k}\|^{2}}$
\STATE$\mathbf{s}_{k+1}=\mathbf{s}_{k}+\boldsymbol{\alpha}_{k}\mathbf{r}_{k}$
\ENDFOR
\ENSURE $\mathbf{\hat{s}}=\mathbf{s}_{K+1}$
\end{algorithmic}
\end{algorithm}

It is easily shown that MINRES minimizes the function $f(\mathbf{s})=\|\mathbf{y}-\mathbf{A}\mathbf{s}\|_{2}^{2}$ in the direction of $\mathbf{r}$. Since MINRES is the simplest RBD algorithm, it requires the filtering matrix $\mathbf{A}$ only to be positive definite. Since the MMSE filtering matrix $\mathbf{A}$ is symmetric positive definite (SPD), the requirement can be met easily. So
\begin{equation}\label{eq:MINRES_vanish}
\begin{aligned}
\|\mathbf{r}_{k+1}\|_{2}^{2}&=(\mathbf{r}_{k}-\boldsymbol{\alpha}_{k}\mathbf{A}\mathbf{r}_{k},\mathbf{r}_{k}-\boldsymbol{\alpha}_{k}\mathbf{A}\mathbf{r}_{k})\\
&=(\mathbf{r}_{k}-\boldsymbol{\alpha}_{k}\mathbf{A}\mathbf{r}_{k},\mathbf{r}_{k})-\boldsymbol{\alpha}_{k}(\mathbf{r}_{k}-\boldsymbol{\alpha}_{k}\mathbf{A}\mathbf{r}_{k},\mathbf{A}\mathbf{r}_{k}).
\end{aligned}
\end{equation}

For the vector $\mathbf{r}_{k}-\boldsymbol{\alpha}_{k}\mathbf{A}\mathbf{r}_{k}$ is orthogonal to search direction $\mathbf{A}\mathbf{r}_{k}$, thus the right side of Eq.~(\ref{eq:MINRES_vanish}) vanishes and therefore
\begin{equation}
\begin{aligned}
\|\mathbf{r}_{k+1}\|_{2}^{2}&=(\mathbf{r}_{k}-\boldsymbol{\alpha}_{k}\mathbf{A}\mathbf{r}_{k},\mathbf{r}_{k})\\
&=(\mathbf{r}_{k},\mathbf{r}_{k})-\boldsymbol{\alpha}_{k}(\mathbf{A}\mathbf{r}_{k},\mathbf{r}_{k})\\
&=\|\mathbf{r}_{k}\|^{2}(1-\frac{(\mathbf{A}\mathbf{r}_{k},\mathbf{r}_{k})}{(\mathbf{r}_{k},\mathbf{r}_{k})}\frac{(\mathbf{A}\mathbf{r}_{k},\mathbf{r}_{k})}{(\mathbf{A}\mathbf{r}_{k},\mathbf{A}\mathbf{r}_{k})})\\
&=\|\mathbf{r}_{k}\|^{2}(1-\frac{(\mathbf{A}\mathbf{r}_{k},\mathbf{r}_{k})^{2}}{(\mathbf{r}_{k},\mathbf{r}_{k})^{2}}\frac{\|\mathbf{r}_{k}\|_{2}^{2}}{\|\mathbf{A}\mathbf{r}_{k}\|_{2}^{2}}).
\end{aligned}
\end{equation}

For the positive definite matrix $\mathbf{A}$,
\begin{equation}
\frac{(\mathbf{A}\mathbf{x},\mathbf{x})}{(\mathbf{x},\mathbf{x})}\geq\lambda_{min}(\mathbf{A}+\mathbf{A}^{T})/2>0.
\end{equation}

Since matrix $\mathbf{A}$ is positive definite, its inversion $\mathbf{A}^{-1}$ is positive definite, too. Similarly, let $\mathbf{t}=\mathbf{A}\mathbf{x}$ then
\begin{equation}
\frac{(\mathbf{A}\mathbf{x},\mathbf{x})}{(\mathbf{A}\mathbf{x},\mathbf{A}\mathbf{x})}=\frac{(\mathbf{t},\mathbf{A}^{-1}\mathbf{t})}{(\mathbf{t},\mathbf{t})}\geq\lambda_{min}(\mathbf{A}^{-1}+\mathbf{A}^{-T})/2>0.
\end{equation}

Finally, let $\mu(\mathbf{A})$ denotes $\lambda_{min}(\mathbf{A}+\mathbf{A}^{T})/2$, then
\begin{equation}
\|\mathbf{r}_{k+1}\|_{2}^{2}\leq(1-\mu(\mathbf{A})\mu(\mathbf{A^{-1}}))\|r_{k}\|_{2}^{2}.
\end{equation}

From the derivation given, residual norm in MINRES algorithm decreases after each iteration, thus the convergence of MINRES can be proven.

\subsection{Generalized Minimal Residual Algorithm}\label{sec:GMRES}
The Generalized Minimal Residual (GMRES) Algorithm is an iterative method to calculate the solution of nonsymmetric system of linear systems \citep{saad2003iterative}. It is the generalized version of MINRES, GMRES inference canceller was proposed in \cite{abdaoui2007gmres,abdaoui2008gmres} first and in this paper, the essence of GMRES will be introduced. Some computation processes to elaborate the computation process of GMRES are supplemented in this paper. As a projection method based on $\boldsymbol{\kappa}=\boldsymbol{\kappa}_{V}$ in which $\mathbf{\kappa}_{V}$ is $V$-th Krylov subspace, GMRES can minimize the residual norm to approximate the exact solution of $\mathbf{A}\mathbf{s}=\mathbf{y}$ by the vector $\mathbf{s}_{k}\in\boldsymbol{\kappa}_{k}$, where
\begin{equation}
\boldsymbol{\kappa}_{V}=span\{\mathbf{y},\mathbf{A}\mathbf{y},\mathbf{A}^{2}\mathbf{y},...,\mathbf{A}^{V-1}\mathbf{y}\}.
\end{equation}

To avoid the linear independence of vectors $\mathbf{y},\mathbf{A}\mathbf{y}$,..., $\mathbf{A}^{V-1}\mathbf{y}$, Arnoldi iteration \cite{voss2004arnoldi} is used to form orthogonal basis $\mathbf{q}_{1},\mathbf{q}_{2},...,\mathbf{q}_{V}$ for $\boldsymbol{\kappa}_{V}$. Thus vector $\mathbf{s}_{V}\in\boldsymbol{\kappa}_{V}$ can be rewritten as $\mathbf{s}=\mathbf{s}_{0}+\mathbf{Q}_{V}\mathbf{p}_{V}$, where $\mathbf{Q}_{V}$ is an $m$-by-$V$ matrix formed by basis $\mathbf{q}_{1}$,$\mathbf{q}_{2}$,$...$,$\mathbf{q}_{V}$.

Meanwhile, a $(V+1)$-by-$V$ upper Hessenberg matrix $\mathbf{\tilde{H}}_{V}$ is produced in the Arnoldi iteration process, where
\begin{equation}
\mathbf{A}\mathbf{Q}_{V}=\mathbf{Q}_{V+1}\mathbf{\tilde{H}}_{V}.
\end{equation}

Thus, the whole GMRES process can be deduced. Define
\begin{equation}
\begin{aligned}
J(\mathbf{p})&=\|\mathbf{y}-\mathbf{A}\mathbf{s}\|_{2}=\|\mathbf{y}-\mathbf{A}(\mathbf{s}_{0}+\mathbf{Q}_{V}\mathbf{p})\|_{2}\\
&=\|\mathbf{r}_{0}-\mathbf{A}\mathbf{Q}_{V}\mathbf{p}\|_{2}\\
&=\|\beta\mathbf{q}_{1}-\mathbf{Q}_{V+1}\mathbf{\tilde{H}}_{V}\mathbf{p}\|_{2}\\
&=\|\mathbf{Q}_{V+1}(\beta\mathbf{e}_{1}-\mathbf{\tilde{H}}_{V}\mathbf{p})\|_{2}.
\end{aligned}
\end{equation}

Since the column-vectors of $\mathbf{Q}_{V+1}$ are orthogonal, it is easy to understand that
\begin{equation}
J(\mathbf{p})=\|\beta\mathbf{e}_{1}-\mathbf{\tilde{H}}_{V}\mathbf{p}\|_{2}.
\end{equation}

With the definition of $J(\mathbf{p})$, GMRES algorithm minimizes it and make the signal approximating $\mathbf{s}_{0}+\boldsymbol{\kappa}_{V}$. After knowing this, GMRES approximation can be denoted by simple equation
\begin{equation}
\mathbf{s}_{V}=\mathbf{s}_{0}+\mathbf{Q}\mathbf{p}_{V},
\end{equation}
where
\begin{equation}
\mathbf{p}_{V}=\mathop{\arg\min}{\|\beta\mathbf{e}_{1}-\mathbf{\tilde{H}}_{V}\mathbf{p}_{V}\|_{2}}.
\end{equation}

Accordingly, the computation process of GMRES algorithm is shown in \emph{Algorithm}~\ref{GMRES}.
\begin{algorithm}
\caption{Generalized Minimal Residual Algorithm}
\label{GMRES}
\begin{algorithmic}[1]
\REQUIRE $\mathbf{A}$ and $\mathbf{\tilde{y}}$\\
$\mathbf{r}_{0}=\mathbf{\tilde{y}}-\mathbf{A}\mathbf{s}_{0}$, $\beta=\|\mathbf{r}_{0}\|_{2}$ and $\mathbf{q}_{1}=\mathbf{r}_{0}/\beta$\\
Define the $(V+1)$-by-$V$ matrix $\mathbf{\tilde{H}}_{V}$. $\mathbf{\tilde{H}}_{V}=\mathbf{0}$
\FOR{$j=1,\ldots,V$}
	\STATE$\mathbf{w}_{j}=\mathbf{A}\mathbf{q}_{j}$
		\FOR{$i=1,\ldots,j$}
		\STATE$\mathbf{\tilde{H}}(i,j)=\mathbf{q}_{i}^{T}\mathbf{w}_{j}$
		\STATE$\mathbf{w}_{j}=\mathbf{w}_{j}-\mathbf{\tilde{H}}(i,j)\mathbf{q}_{i}$
		\ENDFOR
	\STATE$\mathbf{\tilde{H}}(j+1,j)=\|\mathbf{w}_{j}\|_{2}$
	\IF{$\mathbf{\tilde{H}}(j+1,j)=0$}
		\STATE$V=j$ and go to $13$
	\ENDIF
	\STATE$\mathbf{q}_{j+1}=\mathbf{w}_{j}/\mathbf{\tilde{H}}(j+1,j)$
	\STATE$\mathbf{p}_{V}=\mathop{\arg\min}{\|\beta\mathbf{e}_{1}-\mathbf{\tilde{H}}_{V}\mathbf{p}\|_{2}}$
	\STATE$\mathbf{s}_{k}=\mathbf{s}_{k-1}+\mathbf{Q}_{V}\mathbf{p}_{V}$
\ENDFOR
\ENSURE $\mathbf{\hat{s}} =\mathbf{s}_{V}$
\end{algorithmic}
\end{algorithm}

With the information given, M-MIMO detection problems can be solved easily. However, key step of GMRES is step-$12$ in \emph{Algorithm}~\ref{GMRES}, which is not mentioned in \cite{abdaoui2007gmres,abdaoui2008gmres}. To supplement the process of GMRES and make it easier to be understood, Givens rotation to solve this optimization problem is introduced in this paper and can be seen in Appendix~\ref{app:givens}.

For matrix $\mathbf{A}$, $(\mathbf{A}^{T}+\mathbf{A})/2$ is positive definite, then in the $k$-th iteration,
\begin{equation}\label{eq:gmres_convergence}
\|\mathbf{r}_{k}\|\leq(1-\frac{\lambda_{min}^{2}(1/2(\mathbf{A}^{T}+\mathbf{A}))}{\lambda_{max}(\mathbf{A}^{T}\mathbf{A})})^{n/2}\|\mathbf{r}_{0}\|,
\end{equation}
where $\lambda_{min}(\mathbf{M})$ and $\lambda_{max}{\mathbf{M}}$ denote the minimum and maximum eigenvalue of matrix $\mathbf{M}$, respectively.

While in M-MIMO detection scheme, matrix $\mathbf{A}$ is SPD, then Eq.~(\ref{eq:gmres_convergence}) can be deformed to
\begin{equation}\label{eq:gmres_spd}
\|\mathbf{r}_{k}\|\leq(\frac{\tau_{2}(\mathbf{A})^{2}-1}{\tau_{2}(\mathbf{A})^{2}})^{n/2}\|\mathbf{r}_{0}\|,
\end{equation}
where $\tau_{2}(\mathbf{A})$ is the condition number of $\mathbf{A}$.

From Eq.s~(\ref{eq:gmres_convergence}) and~(\ref{eq:gmres_spd}), it can be seen that residual norm of GMRES strictly decreases after iterations, which shows the convergence of it. Synthesizing Arnoldi GMRES algorithm and Givens rotation, the complete GMRES algorithm is a kind of advanced algorithm as a M-MIMO detection method by minimizing the norm of the residual vector.

\subsection{Conjugate Residual Algorithm}\label{sec:CR}
As can be seen in Section~\ref{sec:GMRES}, complete GMRES algorithm needs too many operations and some of them are square root, and even matrix inversion from Givens rotation, which should be avoided in M-MIMO detection scheme. To remedy this and keep the performance of the algorithm for M-MIMO detection, GMRES can be updated to an advanced version.

Consider GMRES is an algorithm for nonsymmetric problem, while M-MIMO detection is solving a SPD problem, some restrictions can be added to GMRES, which makes the GMRES algorithm involving into the proposed conjugate residual (CR) algorithm. Switching nonsymmetric problems to Hermitian problems, CR can lower the computational complexity of GMRES. Being another Krylov subspace iterative method, CR also minimizes the residual vector in each iteration and is feasible in M-MIMO detection. Computation process of CR algorithm is shown in \emph{Algorithm}~\ref{CR}.
\begin{algorithm}
\caption{CR for MMSE detection}
\label{CR}
\begin{algorithmic}[1]
\REQUIRE $\mathbf{A}$ and $\mathbf{\tilde{y}}$\\
$\mathbf{s}_{0}=\mathbf{0}$, $\mathbf{r}_{0}=\mathbf{\tilde{y}}-\mathbf{A}\mathbf{s}_{0}$, $\mathbf{p}_{0}=\mathbf{r}_{0}$\\
$\mathbf{e}_{0}=\mathbf{A}\mathbf{p}_{0}$, $\mathbf{m}_{0}=\mathbf{A}\mathbf{r}_{0}$
\FOR{$k=1,\ldots,K$}
\STATE$\boldsymbol{\alpha}_{k}=\mathbf{r}_{k-1}^{H}\mathbf{m}_{k-1}/\|\mathbf{e}_{k-1}\|^{2}$
\STATE$\mathbf{s}_{k}=\mathbf{s}_{k-1}+\boldsymbol{\alpha}_{k}\mathbf{p}_{k-1}$
\STATE$\mathbf{r}_{k}=\mathbf{r}_{k-1}-\boldsymbol{\alpha}_{k}\mathbf{e}_{k-1}$
\STATE$\mathbf{m}_{k}=\mathbf{A}\mathbf{r}_{k}$
\STATE$\boldsymbol{\beta}_{k}=\mathbf{r}_{k}^{H}\mathbf{m}_{k}/\mathbf{r}_{k-1}^{H}\mathbf{m}_{k-1}$
\STATE$\mathbf{p}_{k}=\mathbf{r}_{k}+\boldsymbol{\beta}_{k}\mathbf{p}_{k-1}$
\STATE$\mathbf{e}_{k}=\mathbf{m}_{k}+\boldsymbol{\beta}_{k}\mathbf{e}_{k-1}$
\ENDFOR
\ENSURE $\mathbf{\hat{s}} =\mathbf{s}_{K}$
\end{algorithmic}
\end{algorithm}

The output $\mathbf{\hat{s}}$ can be proved to support the convergence of the algorithm \cite{fong2012cg}.  For CR on an SPD system,
\begin{eqnarray}
\|\mathbf{s}_{k}\|^{2}-\|\mathbf{s}_{k-1}\|^{2}=2\alpha_{k}\mathbf{s}_{k-1}^{T}\mathbf{p}_{k-1}+\mathbf{p}_{k-1}^{T}\mathbf{p}_{k-1}\geq0.
\end{eqnarray}

Therefore,
\begin{eqnarray}
\|\mathbf{s}_{k}\|\geq\|\mathbf{s}_{k-1}\|.
\end{eqnarray}

Then, final solution can be expressed as $\mathbf{s}_{l}=\mathbf{s}^{*}$,
\begin{equation}
\begin{aligned}
\mathbf{s}_{l}&=\mathbf{s}_{l-1}+\alpha_{l-1}\mathbf{p}_{l-1}\\
&=\cdots\\
&=\mathbf{s}_{k}+\alpha_{k+1}\mathbf{p}_{k}+\cdots +\alpha_{l-1}\mathbf{p}_{l-1}\\
&=\mathbf{s}_{k-1}+\alpha_{k}\mathbf{p}_{k-1}+\alpha_{k+1}\mathbf{p}_{k}+\cdots +\alpha_{l-1}\mathbf{p}_{l-1}.
\end{aligned}
\end{equation}

From the conclusion above, it can be deduced that
\begin{equation}
\resizebox{.92\hsize}{!}{$
 \begin{aligned}
&\|\mathbf{s}_{l}-\mathbf{s}_{k-1}\|^{2}-\|\mathbf{s}_{l}-\mathbf{s}_{k}\|^{2}
\\
&=2\alpha_{k}\mathbf{p}_{k-1}^{T}(\alpha_{k+1}\mathbf{p}_{k}+\cdots+\alpha_{l-1}\mathbf{p}_{l-1})
+\alpha_{k}^{2}\mathbf{p}_{K-1}^{T}\mathbf{p}_{k-1}\geq0.
 \end{aligned}
 $}
\end{equation}

While for the MMSE linear detection problem, linear equation $\mathbf{A}\mathbf{s}=\mathbf{y}$ is to be solved. Thus
\begin{equation}
\resizebox{.92\hsize}{!}{$
\begin{aligned}
&\|\mathbf{s}_{l}-\mathbf{s}_{k-1}\|_{\mathbf{A}}^{2}-\|\mathbf{s}_{l}-\mathbf{s}_{k}\|_{\mathbf{A}}^{2}\\
&=2\alpha_{k}\mathbf{p}_{k-1}^{T}\mathbf{A}(\alpha_{k+1}\mathbf{p}_{k}+\cdots+\alpha_{l-1}\mathbf{p}_{l-1})
+\alpha_{k}^{2}\mathbf{p}_{k-1}^{T}\mathbf{A}\mathbf{p}_{k-1}\\
&=2\alpha_{k}\mathbf{q}_{k-1}^{T}(\alpha_{k+1}\mathbf{p}_{k}+\cdots+\alpha_{l-1}\mathbf{p}_{l-1})
+\alpha_{k}^{2}\mathbf{q}_{k-1}^{T}\mathbf{p}_{k-1}>0.
\end{aligned}
$}
\end{equation}

The derivation above indicates that the residual norm is strictly decreasing. Thus CR is feasible for massive MIMO detection.

\section{Numerical Results and Comparison}\label{sec:Numerical}

\subsection{Results with Different Antenna Configurations}
With $64$-QAM and i.i.d. channel model, the bit-error-rate (BER) comparison between each RBD algorithm and two antenna configurations are considered. Here iteration time $k$ is set as $2$, $3$ and $4$, respectively.

\begin{figure}[htbp]
\centering\includegraphics[width=.9\linewidth]{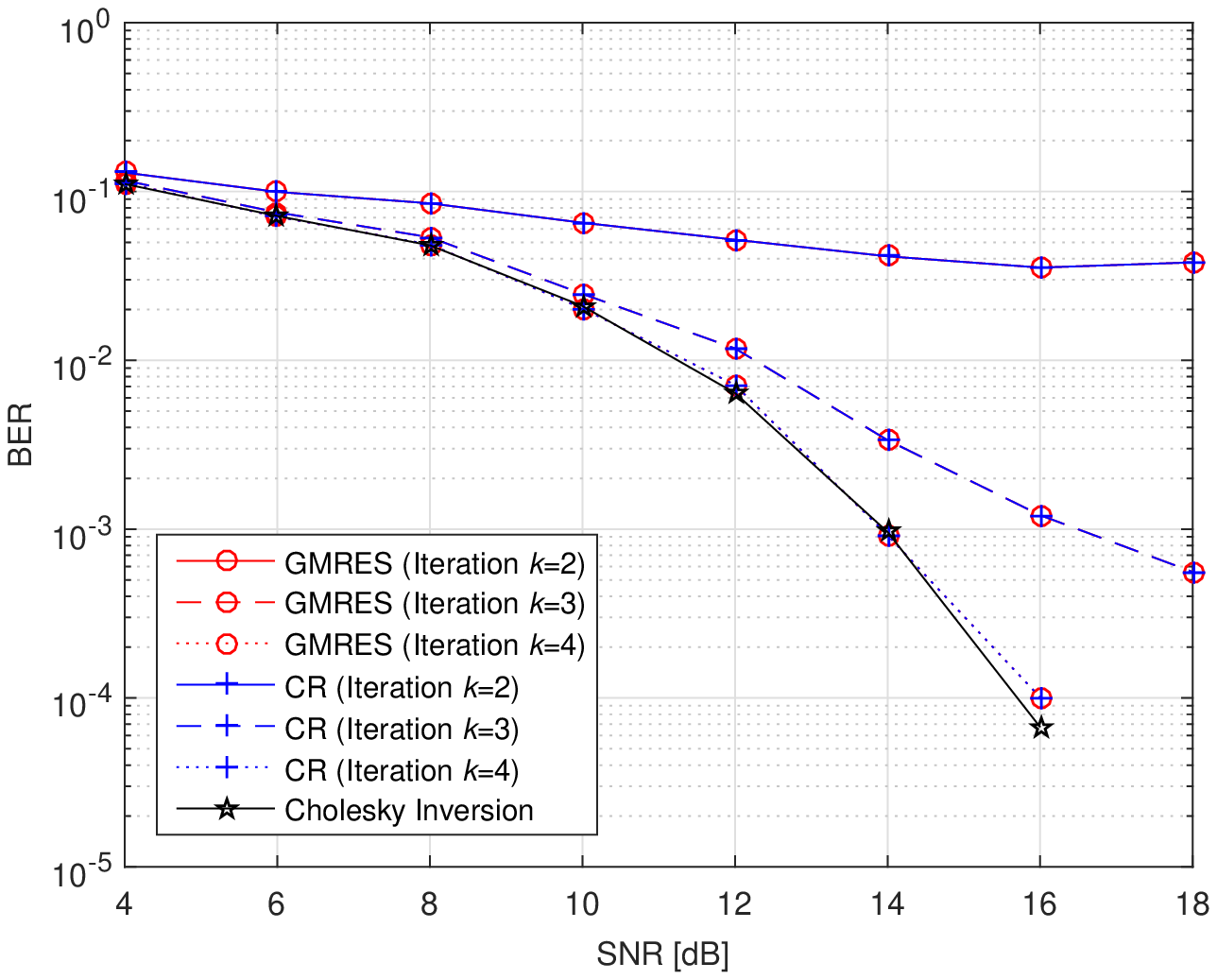}
\caption{Performance comparison with $N\times M=128\times 16$.}\label{fig:gmres_cr_comp}
\end{figure}

It is worth noting that because CR algorithm is a derivation of GMRES algorithm in M-MIMO scheme, they have the same BER performance as mentioned in Section~\ref{sec:CR}. For better elaboration, it is also shown in Fig.~\ref{fig:gmres_cr_comp}. It can be seen that when $k=4$, both of them approximate traditional matrix inversion. To be specific, when BER$=10^{-4}$, CR has only $0.28$ dB gap between \emph{Cholesky} decomposition.

\begin{figure}[htbp]
\centering\includegraphics[width=.9\linewidth]{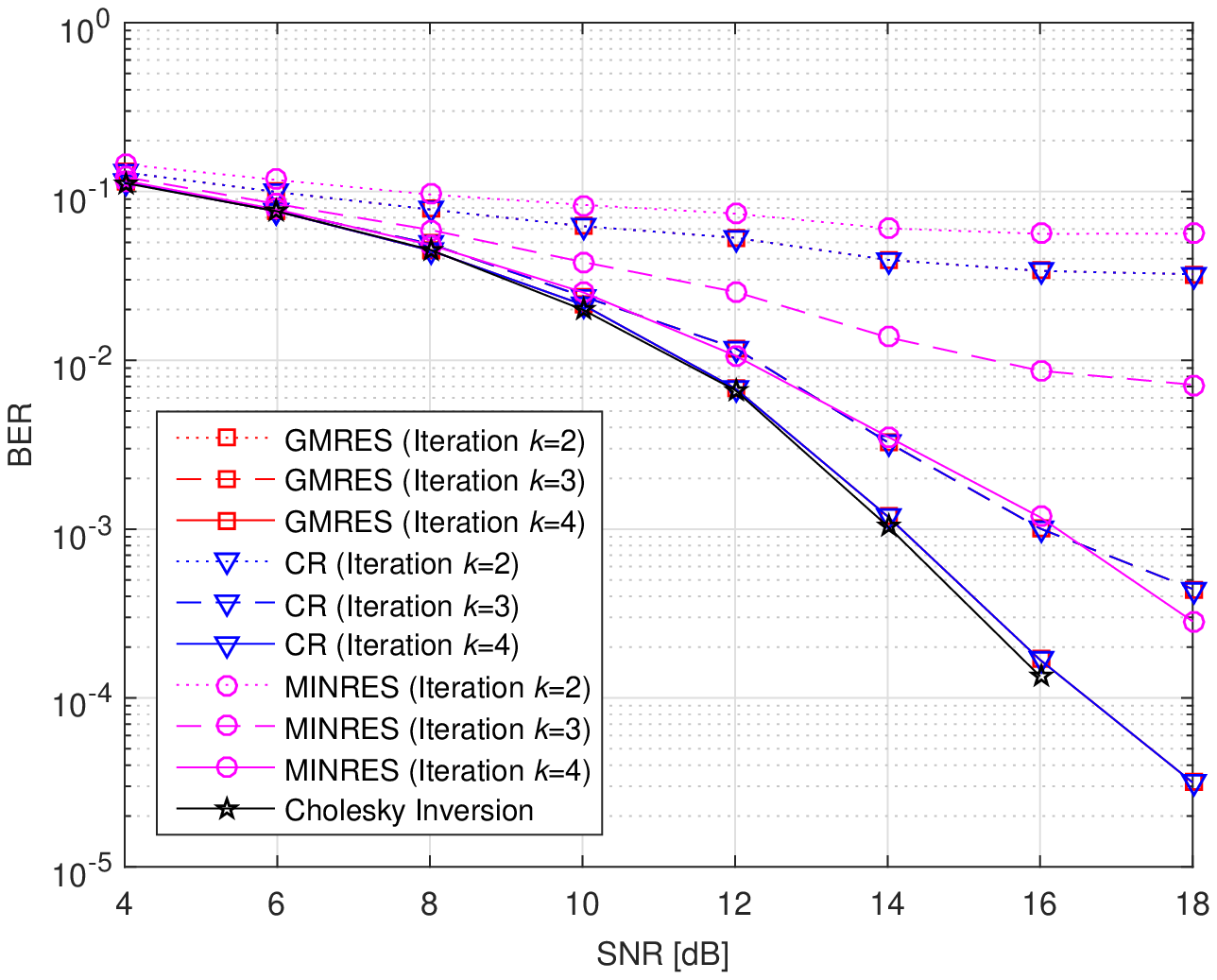}
\caption{Performance comparison with $N\times M=128\times 16$.}\label{fig:128_16}
\end{figure}

\begin{figure}[htbp]
\centering\includegraphics[width=.9\linewidth]{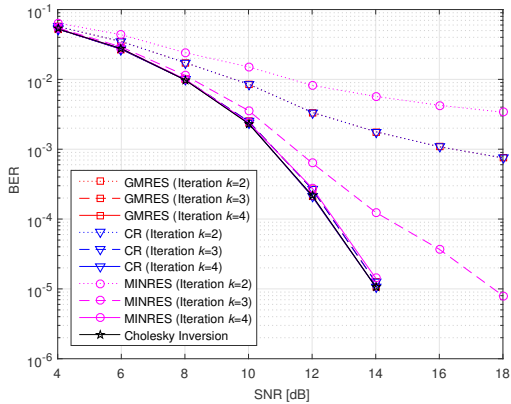}
\caption{Performance comparison with $N\times M=128\times 8$.}\label{fig:128_8}
\end{figure}

Here Fig.s~\ref{fig:128_16} and \ref{fig:128_8} compares BER performance of each RBD algorithm when the antenna configuration is $N\times M=128\times 16$ and $128\times 8$, respectively. In Fig.~\ref{fig:128_16}, when BER$=10^{-3}$ and iteration time $k=4$, MINRES has $2.3$ dB drawback compared with \emph{Cholesky} decomposition while GMRES and CR have only $0.18$ dB gap with \emph{Cholesky} decomposition. Performance of RBD algorithms improve a lot along with the increment of iteration time $k$. Meanwhile, GMRES and CR outperform MINRES a lot. For example, when BER$=10^{-2}$ and iteration time $k=3$, CR and GMRES outperform MINRES by $2.73$ dB SNR gap.

For another antenna configuration $N\times M=128\times 8$, as shown in Fig.~\ref{fig:128_8}, RBD algorithms perform well in approximating \emph{Cholesky} decomposition scheme. MINRES has huge performance improvement as iteration time increases. Take BER$=7\times 10^{-2}$ for instance, MINRES has $6.2$ dB gain when iteration time increases from $k=2$ to $k=3$. CR and GMRES have almost the same performance with exact matrix inversion when iteration time $k\geq 3$, in which condition SNR gap between them is less than $0.2$ dB.

\subsection{Results with Different Correlation Conditions}
\begin{figure}[htbp]
\centering\includegraphics[width=.9\linewidth]{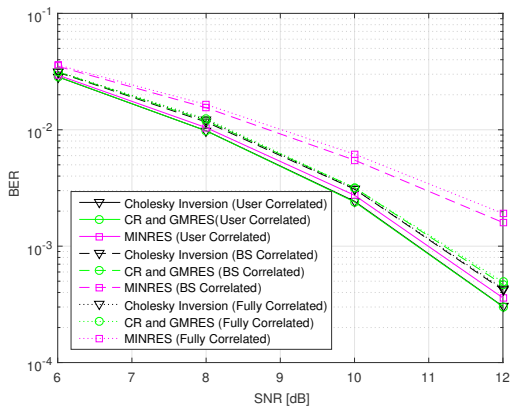}
\caption{Performance comparison with correlations.}\label{fig:correlation}
\end{figure}

Consider $N\times M=128\times 8$ M-MIMO system and iteration time is $4$, BER performances of each RBD algorithm and \emph{Cholesky} decomposition are given in Fig.~\ref{fig:correlation}. Here three conditions are considered: \romannumeral1) \emph{User Correlated} case ($\zeta_{t}=0.2, \zeta_{r}=0$), \romannumeral2) \emph{BS correlated case} ($\zeta_{t}=0, \zeta_{r}=0.3$), \romannumeral3) \emph{Fully Correlated} case: ($\zeta_{t}=0.2, \zeta_{r}=0.3$).

\begin{table*}[ht]
\tabcolsep 2mm
\renewcommand{\arraystretch}{1.5}
\footnotesize
\caption{Complexity Comparison of Different Algorithms.}
\centering
\begin{center}
\begin{tabular}{c || c | c | c }
\hline
\textbf{Operation} & \textbf{MINRES} & \textbf{GMRES} & \textbf{CR}\\
\hline
 \textbf{Addition} & $2kM$ & $(\frac{1}{2}k^{2}+\frac{3}{2}k+1)M$ & $(4K+1)M$ \\
\hline
 \textbf{Multiplication} & $4kM^{2}+2kM$ & $(\frac{5}{2}k^{2}+\frac{1}{2}k+1)M^{2}+(\frac{1}{2}k^{2}+\frac{1}{2}k)M$ & $(k+3)M^{2}+8kM$ \\
\hline
\end{tabular}\label{tab:complexity_comparison}
\end{center}
\end{table*}

In Fig.~\ref{fig:correlation}, as the correlation factor $\zeta$ varies, performance of GMRES and CR remain stable and the performance loss is less than $0.5$ dB. MINRES algorithm will suffer from the change of the correlation condition. However, MINRES loses up to $1.7$ dB when BER$=9\times 10^{-2}$. Thus RBD algorithms are not very sensitive to correlation conditions for M-MIMO.

\section{Computational Complexity Analysis}\label{sec:complexity}
Computational complexity of each RBD algorithm is compared to describe the complexity issue of RBD algorithms. In this section computational complexity is analyzed for better understanding of RBD algorithms. As mentioned in Section~\ref{sec:MINRES}, MINRES algorithm is the basic algorithm in RBD algorithms. GMRES algorithm is the generalized version of MINRES and is complex in computation process. To meet the requirement of M-MIMO system, GMRES can be derived into CR algorithm, which is suitable for M-MIMO detection. Table~\ref{tab:complexity_comparison} concludes the comparison of different algorithms in terms of complex-valued additions and complex-valued multiplications.

\begin{figure}[htbp]
\centering\includegraphics[width=.9\linewidth]{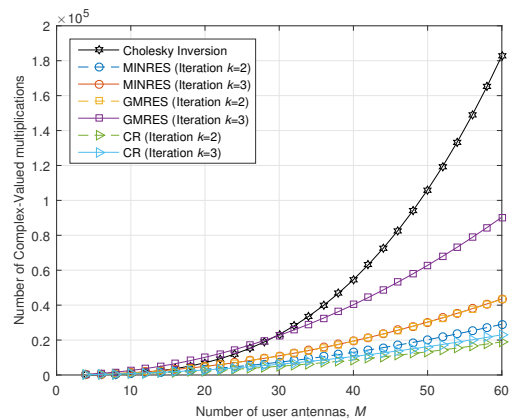}\label{fig:complexity_comparison}
\caption{Computational complexity comparison.}
\end{figure}

The detection complexity is mainly contributed by complex-valued multiplication. The complexity of each algorithm is compared with \emph{Cholesky} decomposition. Suppose the antenna number at BS is $128$ and SNR is $20$ dB. Complexity comparison is shown in Fig.~\ref{fig:complexity_comparison}.

\subsection{Complexity of MINRES}
Being the basic RBD, MINRES has the simplest computation process, though in Fig.~\ref{fig:complexity_comparison} its complexity is not the least. However, when user antenna number is $60$, MINRES can achieve $76\%$ complexity reduction compared with traditional matrix inversion after $3$ iterations. Thus the complexity reduction outweighs the performance loss in terms of trade-off.

\subsection{Complexity of GMRES}
As a generalized version of MINRES, GMRES has more application scenarios. However, its complexity rises also. As shown in Fig.~\ref{fig:complexity_comparison}, GMRES has higher complexity than other RBD algorithms. Similarly, when user antenna number is $60$, GMRES reduces the complexity of traditional method by $50\%$ after $3$ iterations.

\subsection{Complexity of CR}
It is clear that CR has the lowest complexity of RBD algorithms: in the same condition, CR reduces the complexity of traditional method by $87\%$ when user antenna number is $60$ after $3$ iterations. Having the BER performance in Section~\ref{sec:Numerical}, CR is the best algorithm in RBD algorithms and can substitute GMRES in M-MIMO.

\section{Hardware Architecture for RBD Algorithms}\label{sec:hardware}
Computational process of RBD algorithms is introduced in Section~\ref{sec:RBD}. To further elaborate RBD algorithms, corresponding hardware architectures are shown in Section~\ref{sec:hardware_minres} and~\ref{sec:hardware_cr}. Since GMRES algorithm maintains the same BER performance as CR algorithm with unaffordable computational complexity, GMRES algorithm is not hardware friendly. Thus the implementation of GMRES is replaced by CR algorithm. A method to unify the hardware design is also proposed in Section ~\ref{sec:normalized}. Using the new design method, RBD algorithms can be designed by only two basic modules. Unified architectures of MINRES and CR are proposed in Section~\ref{sec:normalized} to validate the design method.

\subsection{Hardware Architecture of MINRES Algorithm}\label{sec:hardware_minres}
\begin{figure}[htbp]
\centering\includegraphics[width=.9\linewidth]{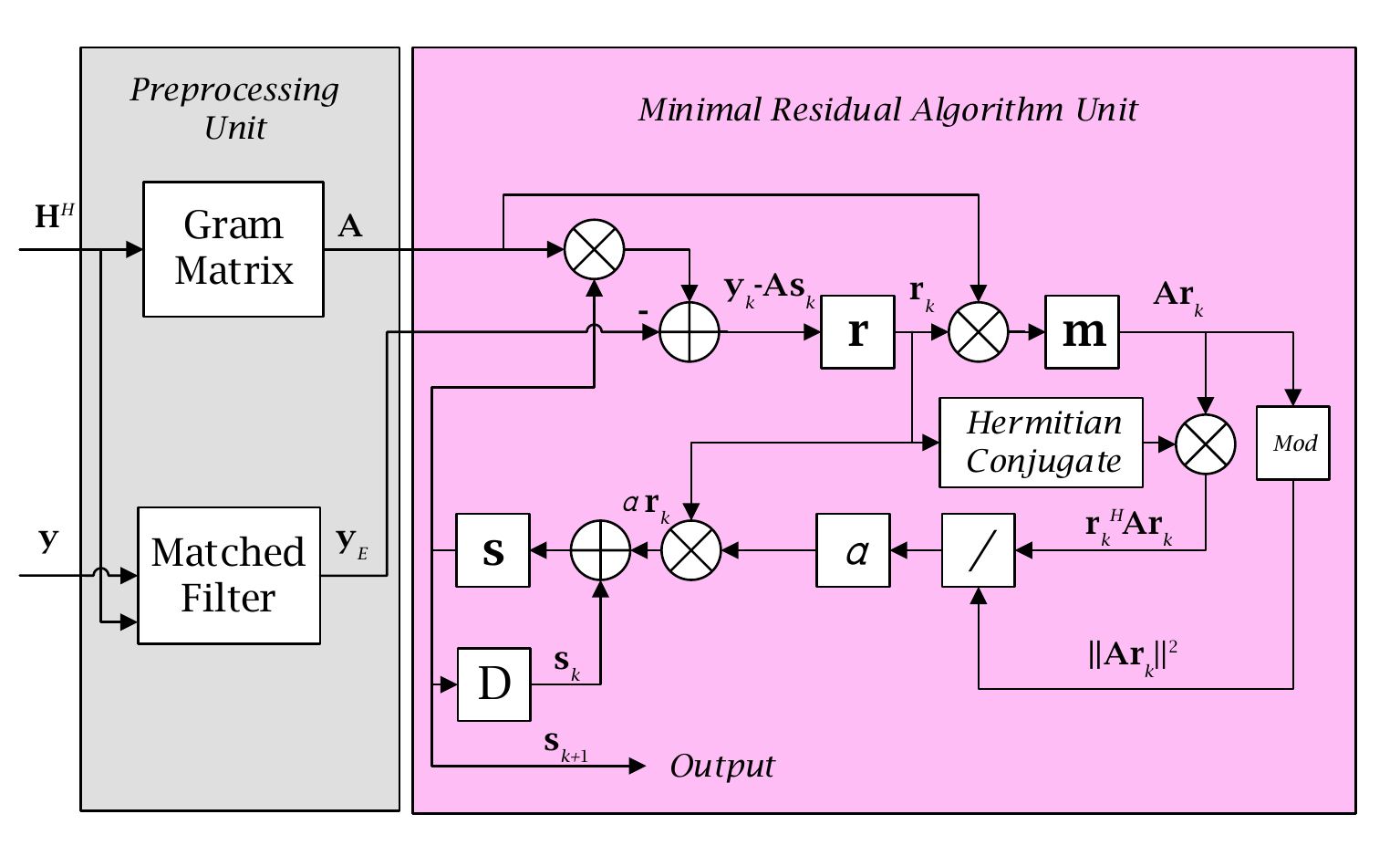}
\caption{Hardware architecture of MINRES method.}\label{hardware:minres}
\end{figure}

As mentioned in Section~\ref{sec:MINRES}, MINRES is the most basic algorithm of RBD algorithm, thus the hardware architecture of it is not very complex. Being divided into two units, Fig.~\ref{hardware:minres} shows the architecture of MINRES algorithm. Preprocessing Unit computes the output of matched filter $\mathbf{\tilde{y}}$ and matrix $\mathbf{A}$. Minimal Residual Algorithm Unit is the main unit of the architecture and it minimizes the residual norm in each iteration. In Fig.~\ref{hardware:minres}, $\mathbf{\tilde{y}}$ is denoted by $\mathbf{y}_{E}$ and the symbol output is denoted by $\mathbf{s}_{k+1}$, where the index $k$ is iteration time.
\begin{figure*}[htbp]
\centering\includegraphics[width=0.6\linewidth]{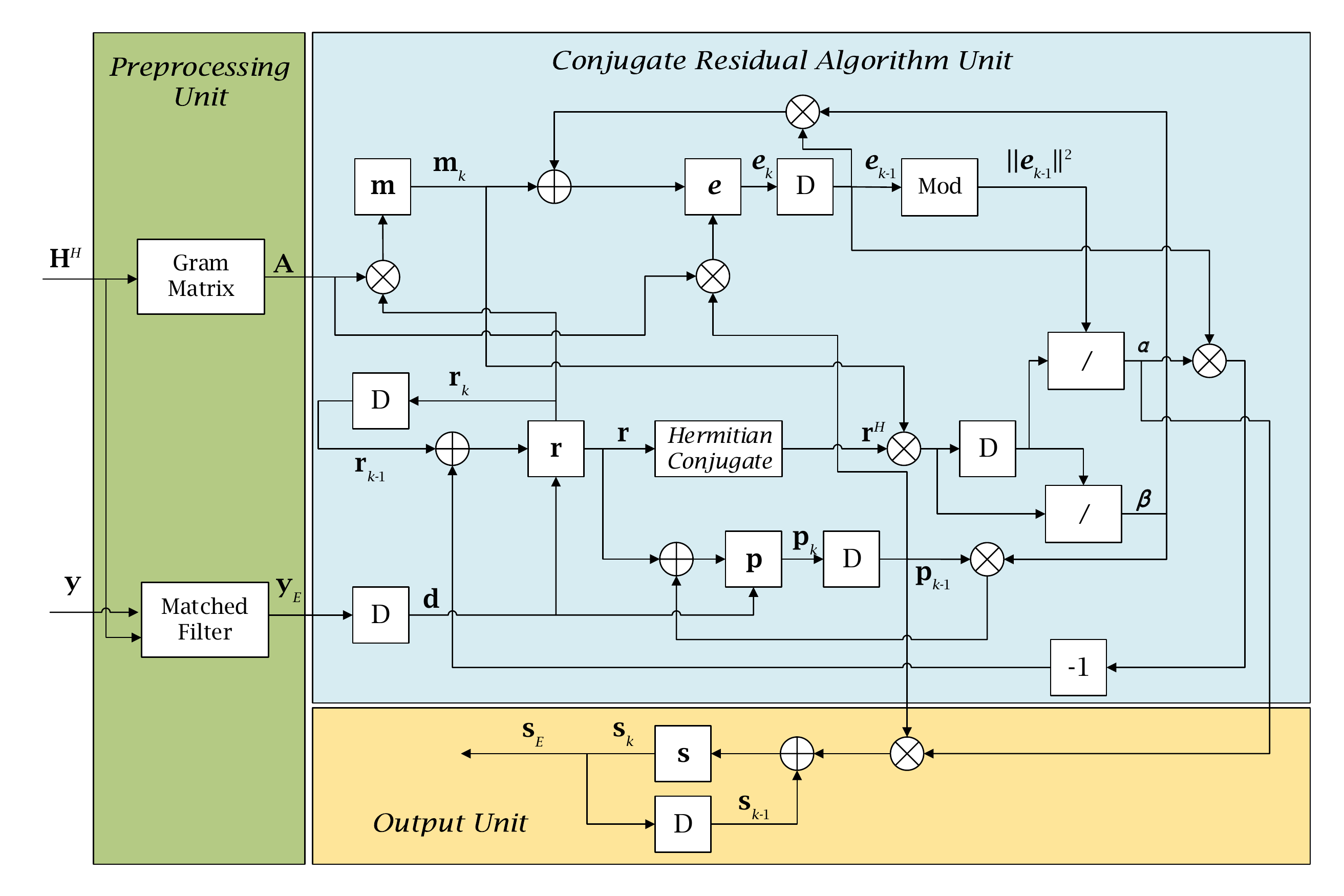}
\caption{Hardware architecture of CR method.}\label{fig:hardware_CR}
\end{figure*}

\subsubsection{Preprocessing Unit}\label{sec:preprocess_unit}
In this unit, matched filter module computes $\mathbf{\tilde{y}}$ by $\mathbf{\tilde{y}}=\mathbf{H}^{H}\mathbf{y}$ while MMSE filtering matrix $\mathbf{A}$ is computed by Gram matrix module. Since matrix A is Hermitian, $M\times M$ lower triangular systolic array is adopted to compute it. Each processing element (PE) performs a multiply-accumulate (MAC) operation with same inputs.

\subsubsection{Minimal Residual Algorithm Unit}\label{sec:MINRES_unit}
In this unit, symbol signals are stored and computed iteratively. Square module with $\mathbf{r}$, $\mathbf{m}$, $\mathbf{s}$, $\boldsymbol{\alpha}$ stores the corresponding signals of each iteration. The hermitian of symbol is given after hermitian conjugate module and module with ``/" means division operation, in which the input from downside is the divisor. Module with ``D" is the delayer which can provide the signal of last iteration for the algorithm. ``Mod" module computes the modulus of the input signal. At the end of the iteration, output $\mathbf{s}_{k}$ is the final symbol output of MINRES.

\subsection{Hardware Architecture of CR Algorithm}\label{sec:hardware_cr}
Being another RBD algorithm, CR has much lower computational complexity compared with traditional exact matrix inversion. As is shown in Section\ref{sec:preprocess_unit}, CR has better performance than MINRES. Thus CR performs well in terms of performance within RBD algorithms. Hardware architecture of CR contains three parts: preprocessing unit, conjugate residual algorithm unit, output unit. Same with that of Section~\ref{sec:preprocess_unit}, preprocessing unit computes $\mathbf{\tilde{y}}$ and $\mathbf{A}$. Conjugate residual algorithm unit is the main unit and it computes symbol signals iteratively to minimize residual norm as well.

\subsubsection{Preprocessing Unit}
Functioning as a preliminary unit, this unit has the same architecture with that in Section~\ref{sec:preprocess_unit}.

\subsubsection{Conjugate Residual Algorithm Unit}
As the main computing unit of CR algorithm, this unit adopts similar functional modules with that in Section~\ref{sec:MINRES_unit}. Differently, with the output of preprocessing unit, CR algorithm needs initialization, which is denoted by the input from the downside of the storage module of vector $\mathbf{r}$, $\mathbf{p}$, $\mathbf{e}$, $\mathbf{m}$. Meanwhile, in this architecture the left input of division module is dividend and the upper input is the divisor of division operation.

\subsubsection{Output Unit}
With the output of CR algorithm unit, we provide the estimation of transmitted signal stored in $\mathbf{s}$. When the iteration ends, final symbol output is denoted by $\mathbf{s}_{E}$.

\subsection{Unified Hardware Architecture}\label{sec:normalized}
Being RBD algorithm, MINRES and CR have different architectures. Thus in terms of implementation they are uncorrelated. Thanks to the special characteristic of RBD algorithm that the minimization of residual norm, RBD algorithms can be designed by a unified design method. In this part a design method to normalize the hardware architecture of RBD algorithm is firstly introduced and then unified architecture of MINRES and CR are given.

\subsubsection{Normalizing Design Method of RBD Algorithms}\label{sec:normalizing_method}
The purpose of the normalizing design method is to make the hardware design of RBD algorithms flexible and reusable, users can switch detectors with existing hardware resources as long as they want to. To meet this purpose, synthesizing the characteristic of RBD algorithms, which is minimizing the residual norm, the normalizing design method is then applied.

\begin{figure}
\centering
\subfigure[Iterative module]{\raisebox{3mm}{\includegraphics[width=0.42\linewidth]{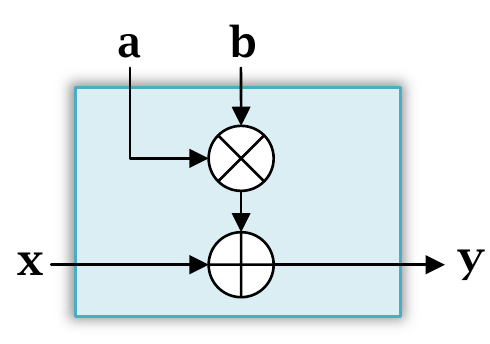}}}
\subfigure[Coefficient module]{\includegraphics[width=0.45\linewidth]{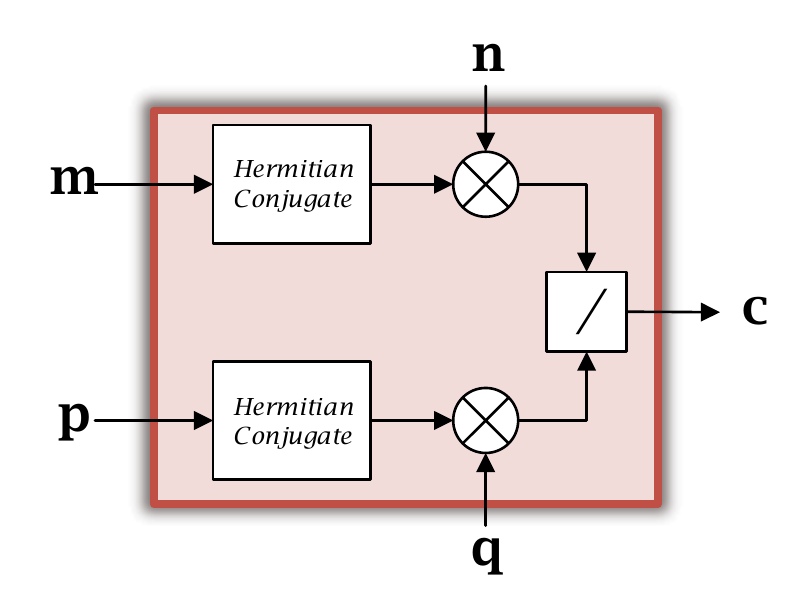}}
\caption{Basic modules of unified hardware architecture.}\label{fig:basic_module}
\end{figure}

Consider the computation process and hardware module of RBD algorithms, this method takes two modules as basic modules: iterative module and coefficient module. Iterative module can iteratively update the signal or compute the residual norm in each iteration. Another module is coefficient module which computes the coefficient of each vector in computation process. Hardware architectures of both basic modules are shown in Fig.~\ref{fig:basic_module}.

Iterative module consists of two operation unit, a multiplier and an accumulator, which performs a MAC operation. Coefficient module consists of two Hermitian conjugate module, a division module and a multiplier, which provides the coefficient for each iteration module. Upper input of this module is the dividend and the input from downside is the divisor. Operation of each module can be denoted by
\begin{equation}
\left\{
\begin{aligned}
\mathbf{y}&=\mathbf{x}+\mathbf{a}\mathbf{b},\\
\mathbf{c}&=\frac{\mathbf{m}^{H}\mathbf{n}}{\mathbf{p}^{H}\mathbf{q}}.
\end{aligned}
\right.
\end{equation}

Having these two basic modules, hardware architectures of RBD algorithms can be unified. Besides basic modules, only some multipliers and delayers are needed. Thus the flexibility of hardware can be improved and the architecture can be reused for further usage. Those two basic modules can also be used in some other iterative detection algorithms. To validate the reasonability of this design method, unified hardware architectures are given in Section~\ref{subsec:nor_MINRES} and ~\ref{subsec:CR}.

\subsubsection{Unified Architecture of MINRES Algorithm}\label{subsec:nor_MINRES}
As the basic algorithm of RBD algorithms, MINRES does not need many basic modules, the unified architecture of it contains two iterative modules and a coefficient module.

Input signal of this unified architecture is also computed from Gram matrix module and matched filter. By normalizing design method, MINRES algorithm adopts two iterative modules to store the residual $\mathbf{r}$ and signal $\mathbf{s}$. Coefficient module serves for the coefficient $\boldsymbol{\alpha}$. Aside from basic modules, unified hardware architecture of MINRES only has an additional multiplier. After the computation of iteration, symbol output is given as the output of an iterative module.
\begin{figure}[htbp]
\centering\includegraphics[width=.9\linewidth]{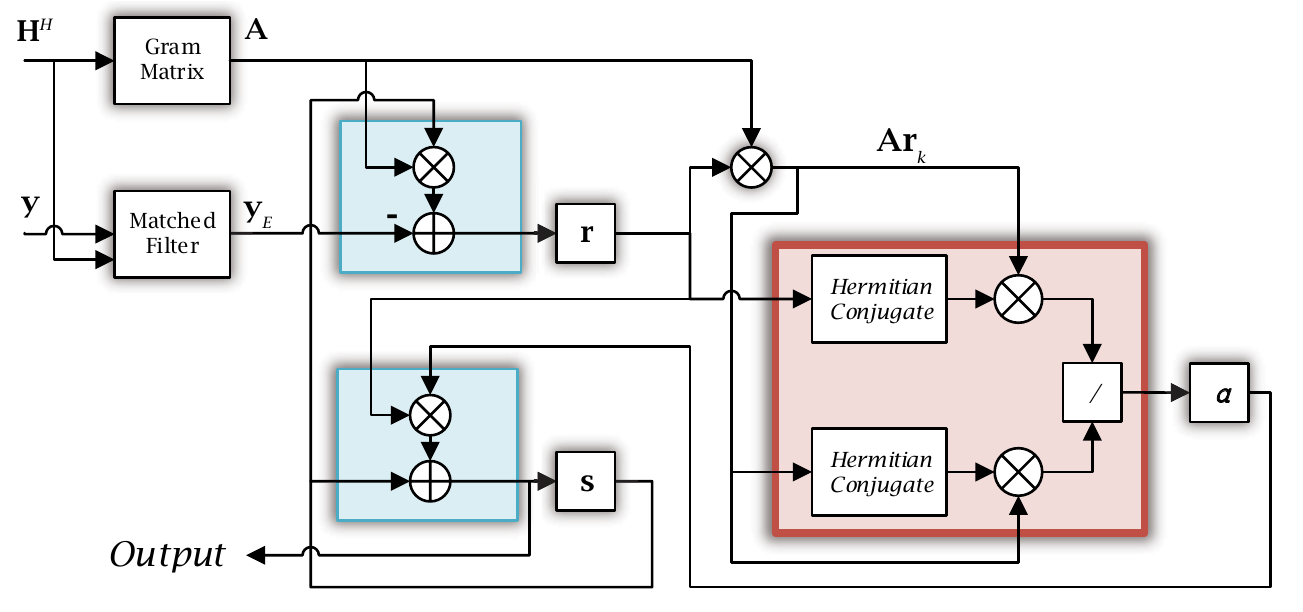}
\caption{Unified architecture of MINRES algorithm.}
\end{figure}

\subsubsection{Unified Hardware Architecture of CR Algorithm}\label{subsec:CR}
\begin{figure*}[htbp]
\centering\includegraphics[width=0.75\linewidth]{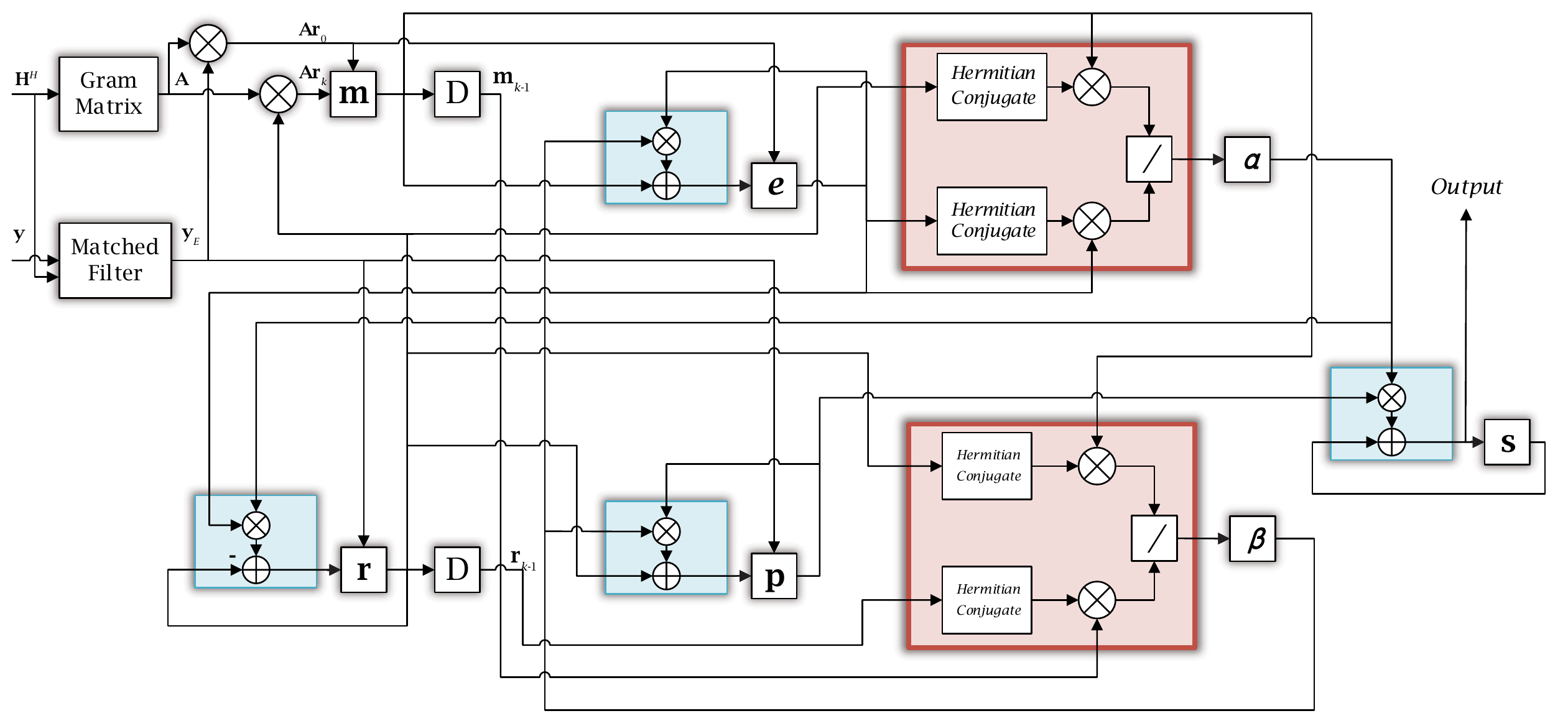}
\caption{Unified architecture of CR algorithm.}\label{fig:nor_hardware_CR}
\end{figure*}

Traditional hardware architecture of CR algorithm is kind of complex as shown in Fig.~\ref{fig:hardware_CR}. After normalization, the architecture is shown in Fig.~\ref{fig:nor_hardware_CR}.

Unified hardware architecture of CR algorithm contains four iterative modules and two coefficient modules. Iterative modules are placed for the storage of signal $\mathbf{r}$, $\mathbf{e}$, $\mathbf{p}$ and $\mathbf{s}$. Coefficient modules store the value of coefficient $\boldsymbol{\alpha}$ and $\boldsymbol{\beta}$. Within each iteration, signal $\mathbf{m}$ is updated by a multiplier and two delayers in the architecture store corresponding signal of last iteration, as mentioned in \emph{Algorithm}~\ref{CR}. Initialization of each signal is the upper input of each module.

With the proposed method in Section ~\ref{sec:normalizing_method}, hardware architectures of RBD algorithms can be unified. Furthermore, the design method can also be applied to other linear iterative detectors like CG.

\section{Conclusion}\label{sec:con}
In this paper, RBD algorithms are first proposed, including MINRES algorithm, GMRES algorithm and CR algorithm. Distinguished from most of other iterative linear detection algorithms, proposed RBD algorithms focus on the minimization of residual norm. Numerical results of different antenna configurations and correlation conditions have demonstrated the approximation to the performance of traditional matrix inversion and the stability of algorithms, respectively. In addition, computational complexity of RBD algorithms are compared and the comparison with matrix inversion shows the complexity reduction advantage of RBD algorithms. Finally hardware architectures of RBD algorithms are first given and the following proposed normalizing design method is adopted, then the unified hardware architectures of RBD algorithms are proposed. Therefore, the proposed RBD algorithms are of good performance, low complexity, and correlation robustness, which are favorable for M-MIMO systems. Future work will be directed towards FPGA implementation of RBD algorithms and further optimization of RBD algorithms.

\begin{acknowledgements}
To be edited.
\end{acknowledgements}

{\appendix
\section{Derivation of Givens rotation}\label{app:givens}
Given the problem $\mathbf{p}=\mathop{\arg\min}{\|\beta\mathbf{e}_{1}-\mathbf{\tilde{H}}_{V}\mathbf{p}\|_{2}}$, knowing that $\mathbf{\tilde{H}}_{V}$ is a $(V+1)$-by-$V$ matrix. It is shown that an over-constrained linear system of $V+1$ equations for $V$ unknowns is given and the minimum can be computed by QR decomposition \cite{wubben2003mmse}. An $(V+1)$-by-$(V+1)$ orthogonal matrix $\mathbf{\Omega}_{V}$ and an $(V+1)$-by-$V$ upper triangular matrix $\mathbf{\tilde{R}}_{V}$ such that $\mathbf{\Omega}_{V}\mathbf{\tilde{H}}_{V}=\mathbf{\tilde{R}}_{V}$.

Because of the characteristic of matrix $\mathbf{\tilde{H}}_{V}$ and $\mathbf{\tilde{R}}_{V}$, they can be denoted as
\begin{equation}
\mathbf{\tilde{H}}_{V+1}=\left[
\begin{array}{cc}
\mathbf{\tilde{H}}_{V} & \mathbf{h}_{V+1}\\
0 & h_{V+2,V+1}
\end{array}
\right],
\mathbf{\tilde{R}}_{V}=\left[
\begin{array}{c}
\mathbf{R}_{V}\\
0
\end{array}
\right],
\end{equation}
where $\mathbf{h}_{V+1}=(h_{1,V+1}$,$...$,$h_{V+1,V+1})^{T}$. Premultiplying the Hessenberg matrix with $\mathbf{\Omega}_{V}$, a nearly triangular matrix can be yielded with zeros and a row with multiplicative identity as

\begin{equation}
\left[
\begin{array}{cc}
\mathbf{\Omega}_{V} & 0\\
0 & 1
\end{array}
\right]
\mathbf{\tilde{H}}_{V+1}=
\left[
\begin{array}{cc}
\mathbf{R}_{V} & \mathbf{r}_{V+1}\\
0 & \rho\\
0 & \sigma
\end{array}
\right].
\end{equation}

If $\sigma=0$, this matrix would be triangular. Givens rotation \cite{ling1991givens} will remedy this as
\begin{equation}
\mathbf{G}_{V+1}=
\left[
\begin{array}{ccc}
\mathbf{I}_{V} &0 & 0\\
0 & c_{V} & b_{V}\\
0 & -b_{V} & c_{V}
\end{array}
\right],
\end{equation}
where
\begin{equation}
c_{V}=\frac{\rho}{\sqrt{\rho^{2}+\sigma^{2}}}\ and\ b_{V}=\frac{\sigma}{\sqrt{\rho^{2}+\sigma^{2}}}.
\end{equation}

After the processing of Givens rotation, matrix $\mathbf{\Omega}_{V}$ can be formed as
\begin{equation}
\mathbf{\Omega}_{V+1}=\mathbf{G}_{V}
\left[
\begin{array}{cc}
\mathbf{\Omega}_{V} & 0\\
0 & 1
\end{array}
\right].
\end{equation}

Meanwhile, a triangular matrix is yielded as
\begin{equation}
\mathbf{\Omega}_{V+1}\mathbf{\tilde{H}}_{V+1}=
\left[
\begin{array}{cc}
\mathbf{R}_{V} & r_{V+1}\\
0 & r_{V+1,V+1}\\
0 & 0
\end{array}
\right],
\end{equation}
where $r_{V+1,V+1}=\sqrt{\rho^{2}+\sigma^{2}}$.

Then given the QR decomposition, the minimization problem can be solved by the transform that
\begin{equation}
\begin{aligned}
\|\mathbf{\tilde{H}}_{V}\mathbf{p}_{V}-\beta\mathbf{e}_{1}\|&=\|\mathbf{\Omega}_{V}(\mathbf{\tilde{H}}_{V}\mathbf{p}_{V}-\beta\mathbf{e}_{1})\|\\
&=\|\mathbf{\tilde{R}}_{V}\mathbf{p}_{V}-\beta\mathbf{\Omega}\mathbf{e}_{1}\|.
\end{aligned}
\end{equation}

Afterwards, using vector $\mathbf{\tilde{g}}_{V}$ to denote $\beta\mathbf{\Omega}\mathbf{e}_{1}$ as
\begin{equation}
\mathbf{\tilde{g}}_{V}=
\left[
\begin{array}{c}
\mathbf{g}_{V}\\
\gamma_{V}
\end{array}
\right],
\end{equation}
where $\mathbf{g}_{V}\in\mathbb{R}_{V}$ and $\gamma_{V}\in\mathbb{R}$.

Finally, norm $\|\mathbf{\tilde{H}}_{V}\mathbf{p}_{V}-\beta\mathbf{e}_{1}\|$ can be denoted by
\begin{equation}
\begin{aligned}
\|\mathbf{\tilde{H}}_{V}\mathbf{p}_{V}-\beta\mathbf{e}_{1}\|&=\|\mathbf{\tilde{R}}_{V}\mathbf{p}_{V}-\beta\mathbf{\Omega}_{V}\mathbf{e}_{1}\|\\
&=\bigg\|
\left[
\begin{array}{c}
\mathbf{R}_{V}\\
\mathbf{0}
\end{array}
\right]
\mathbf{p}_{V}-
\left[
\begin{array}{c}
\mathbf{g}_{V}\\
\gamma_{V}
\end{array}
\right]
\bigg\|.
\end{aligned}
\end{equation}

So vector $\mathbf{p}$ that minimizes the norm is
\begin{equation}
\mathbf{p}_{V}=\mathbf{R}_{V}^{-1}\mathbf{g}_{V},
\end{equation}
where vector $\mathbf{g}_{V}$ can be updated easily and the minimization problem can be solved.

\bibliographystyle{unsrt}
\bibliography{IEEEabrv,rbdbib}

\end{document}